\newcommand{\be}{\begin{equation}}
\newcommand{\ee}{\end{equation}}
\newcommand{\beeq}{\begin{eqnarray}}
\newcommand{\eeeq}{\end{eqnarray}}

\def\funp{{I\!\!P}}
\def\xp{x_{{I\!\!P}}}

\def\alphaem{\alpha_{em}}

\def\rbo{{\bf r}}

\def\Qdash{\overline{Q}}

\documentclass[12pt]{iopart}
\usepackage{iopams,epsfig}

\begin{document}

\title{Saturation and geometric scaling in DIS at small $x$
}

\author{Krzysztof Golec-Biernat\footnote[2]{golec@mail.desy.de}
}

\address{II Institute of Theoretical Physics, Hamburg University, Germany}

\address{Institute of Nuclear Physics, Krak\'ow, Poland}

\begin{abstract}
We present various aspects of the saturation model which provides good description of
inclusive and diffractive DIS at small $x$. The model uses parton saturation ideas
to take into account unitarity requirements. A new scaling
predicted by the model in the small $x$ domain is successfully
confronted with the data.

\end{abstract}




\section{Introduction}
\label{sec:1}
Deeply inelastic electron-proton scattering   (DIS)
at small value of the Bjorken variable $x$ ($\ll 1$) attracted  a lot of attention,
mostly due to the experimental results from HERA. From the theory side,
the small-$x$  DIS opens a new
kinematic regime for the QCD studies. The predicted by QCD, and confirmed by the data
\cite{F2RISE},
strong rise of the proton structure functions with decreasing
$x$ is the indication that at small $x$ the proton  structure is dominated by dense gluon
systems. In DIS at moderate values of $x$, the linear QCD evolution equations
lead to good description of this process, explaining scaling violation. At small $x$,
however, the problem is more complicated since recombination processes between
partons in a dense system have to be taken into account. At the formal level,
these processes allow to restore unitarity of the DIS cross sections as  $x\rightarrow 0$,
violated in the QCD description based on the linear evolution equations,
by taming the rise of the parton distributions.
This effect is called {\it parton saturation} \cite{GLR}.

It was unclear for a long time whether recombination processes
(or unitarization procedures in general) are important in the HERA kinematic range.
Most of the analyses of this problem
were  based on the results obtained in the double logarithmic
approximation in which $x$ is small but $Q^2$ is large. The linear evolution equations
in this case are modified by  non-linear terms describing parton recombination.
In such approximation it was found that
the impact of non-linearity is either small or masked by
the choice of initial conditions for the linear evolution \cite{SATHER}.

The breakthrough  came from realization that  the proper starting point
is the high energy QCD factorization formula for the structure functions,
derived under the assumption that
$x$ is small but $Q^2$ is arbitrary (provided $Q^2\gg \Lambda^2$ and
pQCD is justified). This allows to find the necessity for unitarity,
realized in the spirit of parton saturation, in the transition
from DIS to small $Q^2$ scattering \cite{GBW1}. The full potential of the unitary
description is revealed for diffractive DIS \cite{GBW2}. In particular,
the constant ratio of diffractive to total DIS cross sections
as a function of $x$
and $Q^2$ is naturally explained.

The approach  we are going to present is partially phenomenological.
We take into account
several important features of parton saturation
in the parameterization
of the QCD interactions in a dense gluon system,
especially the existence of a saturation scale. The proposed
parameterization is  very efficient in the description of the data and may serve
as a guidance for the detailed QCD studies.

\section{Saturation model}
\label{sec:2}

\begin{figure}[b]
  \vspace*{-0.8cm}
     \centerline{
         \epsfig{figure=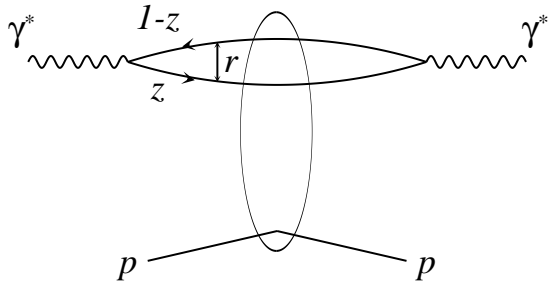,width=9cm}
           }
\vspace*{-0.3cm}
\caption{\it Schematic representation of the basic factorization
in inclusive DIS at small~$x$.
\label{fig:0}}
\end{figure}

The saturation  model was formulated and compared at length to DIS data in
\cite{GBW1,GBW2}. Here we describe this  model, discussing some details which
were not presented in the original formulation. For related
approaches see \cite{OTHERS}.

In the rest frame of the proton, the QCD description of
DIS at small $x$ can be
interpreted as a two-step process. The virtual photon (emitted
by the incident electron) splits into a $q\bar{q}$ dipole which
subsequently interacts with the proton. The proton structure function $F_2$
results from the high energy  QCD factorization theorem. In terms
of  virtual photon-proton cross sections  $\sigma_{T,L}$
for the transverse and longitudinal polarized photons  \cite{NIKO1}
\be
\label{eq:1}
F_2(x,Q^2)\,=\,{Q^2}/{4\pi^2\alphaem} \,(\sigma_T\,+\,\sigma_L)
\ee
and
\be
\label{eq:2}
\sigma_{T,L}\,=\,  \int d^2{\bf{r}}\, dz\, |\Psi_{T,L}(\rbo,z,Q^2)|^2\,\, \hat{\sigma}(x,r),
\ee
where $\Psi_{T,L}$ is the light-cone {\it wave function} of the virtual photon
and $\hat{\sigma}$ is the {\it dipole cross section}
describing  the interaction of the $q\bar{q}$  dipole with the proton.  In
equation (\ref{eq:2})
$\bf{r}$ is the transverse separation of the $q\bar{q}$ pair
and $z$ is the photon's momentum fraction carried by the quark, see Figure \ref{fig:0}.
Thus $(\rbo,z)$ are good quantum numbers
conserved by the interaction in the considered approximation.

The photon wave functions are known from pQCD,
e.g. for transverse photons and massless quarks
\be
\label{eq:3}
|\Psi_{T}(\rbo,z,Q^2)|^2
\,=\,
\frac{3\,\alpha_{em}}{2\pi^2}\, \sum_f e_f^2\,
[z^2+(1-z)^2]\ \Qdash^2\, K_1^2(\Qdash r),
\ee
where $\Qdash^2=z(1-z)\, Q^2$  and  $K_1$ is the Bessel function.
The computation of the dipole cross section $\hat{\sigma}$ has been attempted within
pQCD
assuming different types of net colourless gluon exchange (e.g. DGLAP or
BFKL ladders). Most of these attempts, however,  are
plagued  by problems with unitarity of finally computed cross
sections.

\begin{figure}[ht]
\vspace*{-1cm}
\begin{center}
\mbox{
\epsfig{figure=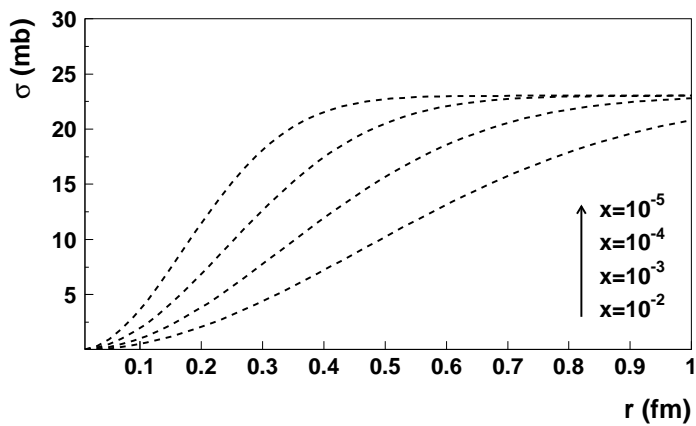,width=11cm}
}
\end{center}
\vspace*{-1cm}
\caption{Dipole cross section for different values of $x$.
\label{fig:1}}
\end{figure}

In our approach
we built in unitarity in the dipole
cross section by proposing the following phenomenological form
\be
\label{eq:4}
\hat{\sigma}(x,r) = \sigma_0\,
\left\{1-\exp\left(-\frac{r^2}{4 R_0^2(x)}\right)\right\},
\ee
with $R_0(x)$, called saturation radius,  given by
\be
\label{eq:5}
R_0^2(x) = \frac{1}{Q_0^2}\,\left(\frac{x}{x_0}\right)^{\lambda},
\ee
where
$Q_0^2=1~\mbox{\rm GeV}^2$. The parameters
$\sigma_0=23~\mbox{\rm mb}, x_0=3\cdot 10^{-4}$ and $\lambda=0.29$  were found from the
fit to all inclusive DIS data at $x<0.01$.
At small $r$,  $\hat{\sigma}$ features colour transparency,
$\hat\sigma\sim r^2$, which is purely pQCD phenomenon.
For  large $r$, saturation occurs, $\hat\sigma\simeq\sigma_0$.
The fact that $\hat\sigma$ is limited by the energy independent cross section may be regarded
as a unitarity bound.
The transition between the two regimes is governed by $R_0(x)$.
As illustrated in Figure~\ref{fig:1}, for $x\rightarrow 0$ the transition occurs for decreasing
transverse sizes. This is an essential feature of parton saturation pioneered by
the analysis \cite{GLR} in which internal saturation scale $Q_s(x)$ appears.
We built in such a scale into our model: $Q_s(x)\sim 1/R_0(x)$.

\begin{figure}[t]
\vspace*{-0.5cm}
\begin{center}
\mbox{
\epsfig{figure=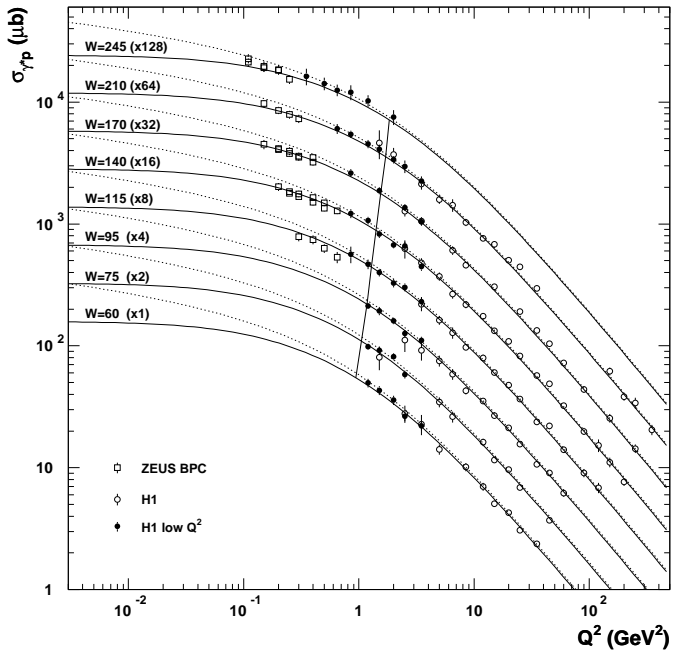,width=10cm}
           }
\end{center}
\vspace*{-0.5cm}
\caption{$\sigma_{\gamma^*p}=\sigma_T+\sigma_L$ for different energies $W$ of the
$\gamma^*p$ system. The solid
lines show the fit results with a light quark mass $m_f=140$ MeV  while  the dotted
lines  with $m_f=0$. The line across the curves indicates the position of the critical line.
\label{fig:2}}
\end{figure}

With the proposed dipole cross section we achieved good description of the DIS data
at small $x$,
including the transition to small $Q^2$ values, see Figure~\ref{fig:2}.
The reason for this can be easily understood
after performing the following qualitative analysis.

\section{Qualitative analysis}
\label{sec:3}

We concentrate on the transverse cross section $\sigma_T$ which dominates in $F_2$.
The dominant contribution to  $\sigma_T$  results from the behaviour of
the Bessel function in (\ref{eq:2}):  $K_1(x)=1/x$ for $x\ll 1$
while for  $x\gg 1$,   $K_1(x)$ is exponentially suppressed.
Thus using (\ref{eq:3}),
\be
\label{eq:6}
\sigma_T\, \sim\, \int_0^\infty \frac{dr^2}{r^2} \int_0^1 dz\
[z^2+(1-z)^2]\ {\hat\sigma(x,r)}\
\Theta\!\left[z(1-z)\, Q^2 r^2 < 1\right],
\ee
where $\Theta(x<1)$ equals 1 if
$x<1$, or 0 otherwise. For $0\le r\le 2/Q$,  the theta function
does not impose any restriction on $z$ and
the $z$-integration gives a constant. Such
a configuration, called {\it symmetric},  is rather
uniform in $z$ with the mean value $1/2$.
For large transverse separations, $r\gg 2/Q$,  the theta function heavily restricts
$z$ to small values:  $z < 1/(Q^2 r^2)$. Now,
the $z$-integration gives the factor $2/(Q^2 r^2)$. In this configuration, called
{\it aligned jet}, $z$ or $(1-z)\approx 0$. Thus, one of
the quarks follows the photon direction  while the other stays with the
proton.

Finally, we obtain
\be
\label{eq:7}
\sigma_T \; \sim \;
\underbrace{
\int_0^{4/Q^2}
\frac{dr^2}{r^2}\ \hat\sigma(x,r)
}_{symmetric}
\;\;+\;\;
\underbrace{
\int_{4/Q^2}^\infty
\frac{dr^2}{r^2}\
 \left(
 \frac{1}{Q^2 r^2}
\right)\ \hat\sigma(x,r)
}_{aligned~ jet},
\ee
where we neglected multiplicative numerical factors.
For the forthcoming analysis let us approximate (\ref{eq:6}) by
\be
\label{eq:8}
\hat\sigma(x,r)\;=\;
\left\{ \begin{array}{ll}
\sigma_0\,r^2/4R_0^2(x)~~~~~  &     \mbox{\rm for~~~ $r\le 2 R_0(x)$} \\ \\
\sigma_0~~~~~              &     \mbox{\rm for~~~ $r> 2 R_0(x)$}\,,
                     \end{array}
             \right.
\ee
which form contains all essential features of the exact
formula. The leading $Q^2$-behaviour  of $\sigma_T$   depends on the relation
between two scales:
the {\it characteristic size} of the $q\bar{q}$ dipole  $1/Q$ and the
{\it mean transverse distance}
between partons given by $R_0(x)$.

If the partonic system is dilute $1/Q\ll R_0(x)$, and from (\ref{eq:7},\ref{eq:8}) we find
\beeq
\label{eq:9}
\sigma_T \;\sim \;
\underbrace{\frac{\sigma_0}{Q^2 R_0^2}}_{r<2/Q}
~\;+\;~
\underbrace{\frac{\sigma_0}{Q^2 R_0^2}\,\ln\, (Q^2 R_0^2)}_{2/Q<r<2R_0}
~\;+\;~
\underbrace{\frac{\sigma_0}{Q^2 R_0^2}}_{r>2R_0}\,,
\eeeq
where we indicated the $r$-integration regions.
In terms of the structure function (\ref{eq:1})   we obtain scaling with logarithmic
violation. Notice that  all dipole sizes, including those from the non-perturbative region,
contribute to scaling. The assumption that $R_0^2\sim x^{\lambda}$ leads to the power-like
behaviour, $F_2\sim x^{-\lambda}$, observed in the data.

When the system of partons
becomes dense for the dipole probe and $1/Q\gg R_0(x)$, a different behaviour is found
\beeq
\label{eq:10}
\sigma_T \;\sim \;
\underbrace{\sigma_0}_{r<2R_0}
\;+\;~
\underbrace{\sigma_0\,\ln\left(\frac{1}{Q^2 R_0^2}\right)}_{2R_0<r<2/Q}
~\;+\;
\underbrace{\sigma_0}_{r>2/Q}\,.
\eeeq
Now, the structure function is in agreement
with unitarity:  $F_2\sim Q^2\,\sigma_0\,\ln(1/x)$.
The change of the behaviour of $\sigma_T$ from (\ref{eq:9}) to (\ref{eq:10})
when $Q^2$ increases while $W^2$ stays fixed
($x=Q^2/W^2\rightarrow 0$) is shown in Figure~\ref{fig:2}.    The same transition
is obtained in the
Regge limit of DIS,  $Q^2$ fixed and $W^2\rightarrow \infty$, since the $\gamma^*p$
center-of-mass energy
$W$ enters the description only through the Bjorken variable $x$.

The transition between the scaling and unitary behaviour of the structure function
depends on $x$ and occurs in the region of the $(x,Q^2)$-plane
marked by the {\it critical line}: $1/Q=R_0(x)$. In our interpretation, at the critical
line
the characteristic size of the dipole probe equals the mean transverse separation
between partons.
The characteristic feature of parton saturation is that the transition occurs
at increasing values of  $Q^2$ when $x\rightarrow 0$.
We found that at HERA the transition region is situated for  $Q^2\approx 1-2~\mbox{\rm GeV}^2$,
which justifies the use of pQCD.

\section{Photoproduction limit}
\label{sec:4}

\begin{figure}
  \vspace*{-0.5cm}
     \centerline{
         \epsfig{figure=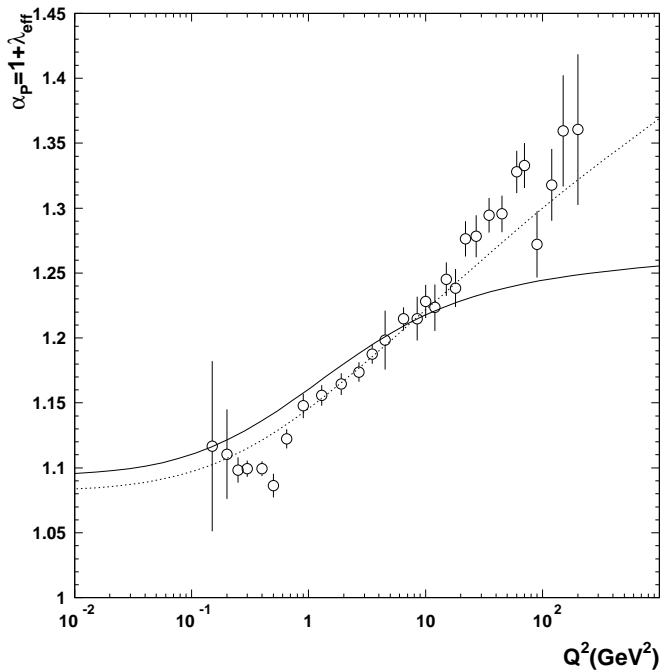,width=10cm}
           }
\vspace*{-0.3cm}
\caption{The  effective pomeron intercept    $\alpha_P=1+\lambda$ as a function of $Q^2$,
found from the dependence: $\sigma_{\gamma*p}\sim ({W^2})^{\lambda}$.  The solid line
corresponds to the saturation model (\ref{eq:4}) while the dashed line  shows the
effect when the model is modified to include the DGLAP evolution (Section~\ref{sec:6}).
The data points are from ZEUS.
\label{fig:3}}
\end{figure}

It is interesting to consider a formal limit $Q^2\rightarrow 0$
in the saturation model. The analyzed cross sections are divergent
in this limit if $m_f=0$. However, if a non-zero quark mass
is assumed, the limit can be performed and we find for
$m_f^2\gg Q^2 \rightarrow 0$,
\beeq
\label{eq:11}
\sigma_T\, &\sim&\, {\sigma_0}\,
\ln\!\left(\frac{1}{m_f^2R_0^2(x)}\right)\,,
\\ \nonumber
\\
\label{eq:12}
\sigma_L\, &\sim&\,  {\sigma_0}\, \frac{Q^2}{m_f^2}\,,
\eeeq
\noindent  where we additionally modify the Bjorken variable formula to allow for
the photoproduction limit
\be
\label{eq:13}
x\;=\; \frac{Q^2+4m_f^2}{W^2}\;.
\ee
\noindent  As expected, the longitudinal cross section vanishes when $Q^2=0$.
We also see that $m_f$ plays a crucial role for the value of the transverse
cross sections. In our analysis
we set $m_f=140~\mbox{\rm MeV}$ to obtain good agreement with the
HERA photoproduction data.
For $Q^2\gg m_f^2$, the light quark mass does not play a significant role.

From the discussion in Section~\ref{sec:3} we know that the energy dependence of
$\sigma_T$ changes from $\sigma_T\sim (W^2)^\lambda$ for large $Q^2$ to
$\sigma_T\sim \ln(W^2)$
in the photoproduction limit. It appears that for each $Q^2$
we can effectively parameterize the energy dependence through the power-like behaviour:
$\sigma_T\sim (W^2)^{\alpha_P(Q^2)-1}$. The found dependence  $\alpha_P(Q^2)$  is shown
in Figure~\ref{fig:3}. Interestingly, $\alpha_P(Q^2)$ interpolates between the soft
and hard pomeron intercept values.

\section{Geometric scaling}
\label{sec:5}

\begin{figure}[b]
  \vspace*{-0.0cm}
     \centerline{
         \epsfig{figure=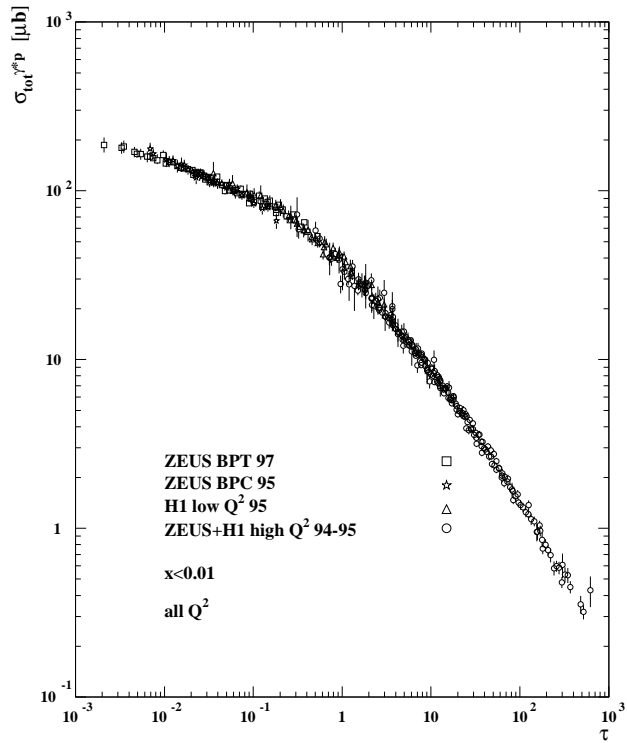,width=8.5cm}
           }
\vspace*{-0.5cm}
\caption{\it Experimental data on $\sigma_{\gamma^* p}$ from the region
$x<0.01$ plotted versus the scaling variable $\tau=Q^2 R_0^2(x)$.
$Q^2$ values are between $0.045$ and $450~\mbox{\rm GeV}^2$.
\label{fig:4}}
\end{figure}

Let us realize that the dipole cross section (\ref{eq:4}) is a function of the
combination $r/R_0(x)$ instead of $x$ and $r$ separately,
\be
\label{eq:14}
\hat\sigma(x,r)\,=\, \hat\sigma(r/R_0(x)).
\ee
 This has profound
consequences for the total cross section
$\sigma_{\gamma^* p}=\sigma_T+\sigma_L$, if the quark mass
$m_f$ in the photon wave functions is neglected. Rescaling the integration
variable  $r\rightarrow r/R_0$ in (\ref{eq:2}), we find that
$\sigma_{\gamma^* p}$ becomes a function of the dimensionless variable
$\tau=Q^2\, R_0^2(x)$, being the ratio of the two geometric scales discussed
in Section~\ref{sec:3},
\be
\label{eq:15}
\sigma_{\gamma^* p}(x,Q^2)\,=\,\sigma_{\gamma^* p}\!\left(Q^2 R_0^2(x)\right)\,.
\ee
The non-zero light quark mass, introduced
to extrapolate the model down to the photoproduction region,
does not lead to a significant
breaking of a new scaling (which we call geometric scaling)
in the small $x$ domain \cite{SGK}.
So does the charm contribution, discussed in detail in \cite{GBW1}.
In Figure~\ref{fig:4}, reproduced from \cite{SGK}, we illustrate geometric  scaling
for the small-$x$ data in a broad range of $Q^2$.

In its essence,
geometric scaling is a manifestation of the presence of  the internal
saturation scale
characterizing  a dense partonic system, $Q_s(x) \sim 1/R_0(x)$.
The presence of such  scale
emerges from a pioneering work  of \cite{GLR}, which was
subsequently analyzed and generalized in
\cite{SATURATION}.
In reference \cite{KOVCH}
the scaling property similar to that postulated in the dipole cross section (\ref{eq:4})
was found and analyzed in detail in \cite{LEVIN}.

\section{DGLAP evolution in the saturation model}
\label{sec:6}

In the limit of large $Q^2$ values, the structure function (\ref{eq:1})
is dominated by the small size transverse contribution. In this limit,
the high energy formula (\ref{eq:1}) should make a contact with the
DGLAP formula for $F_2$. This allows to find the following relation
between the dipole cross section   at small $r$
and the ordinary gluon distribution \cite{GLUON},
\be
\label{eq:16}
\hat\sigma(x,r)\,\approx\,\frac{\pi^2}{3}\, r^2\, \alpha_s\,xg(x,C/r^2).
\ee
The gluon distribution $g(x,\mu^2)$ obeys the DGLAP
evolution equations. The parameter $C$ in (\ref{eq:16}), as well as the argument
of $\alpha_s$, cannot be determined in the considered leading $\log Q^2$ approximation.

In the saturation model (\ref{eq:4}), $\hat\sigma\sim r^2/R_0^2(x)$ at small $r$.
Thus the logarithmic dependence on $r$, resulting from the DGLAP evolution of the gluon,
is not included in the model.
In our recent analysis \cite{BGK}  we include the DGLAP evolution by proposing the following
modification of the saturation model
\be
\label{eq:17}
 \hat\sigma (x,r)\,=\,\sigma_0\,\left\{
1\,-\,\exp\left(-\frac{\pi^2\,r^2\,\alpha_s(\mu^2)\,xg(x,\mu^2)}
{3\,\sigma_0}\right) \right\}\,,
\ee
where the scale $\mu^2\,=\,{C}/{r^2}\,+\,\mu_0^2$,
and the parameters $C$ and $\mu_0^2$ are determined from the best fit
to the DIS data. Additionally, two parameters
of the initial gluon distribution: $xg=A_g\, x^{-\lambda_g}$ are fitted.
The result of the comparison with  the data is shown
in Figure~\ref{fig:3} as the dotted line.
As expected, the proper DGLAP limit
of $\hat\sigma$ significantly improves agreement at large values of $Q^2$ without
affecting the physics of saturation responsible for the transition to small
$Q^2$.
The full discussion of the results will be presented in \cite{BGK}.

\section{Saturation and DIS diffraction}
\label{sec:7}

\begin{figure}[t]
  \vspace*{-1cm}
     \centerline{
         \epsfig{figure=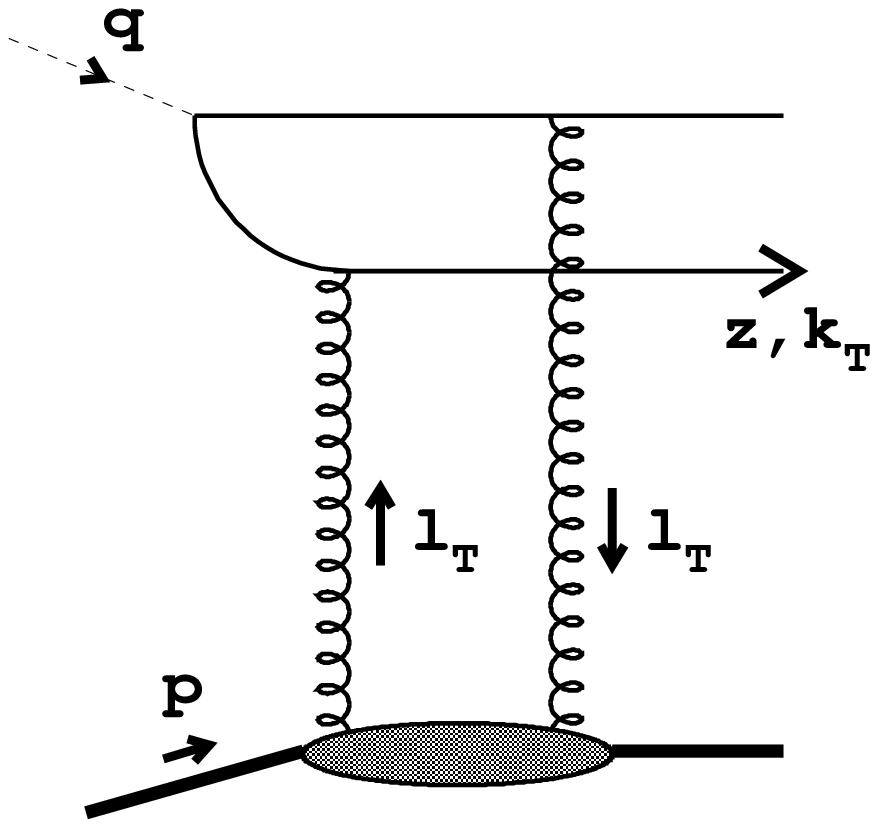,width=6cm}
         \epsfig{figure=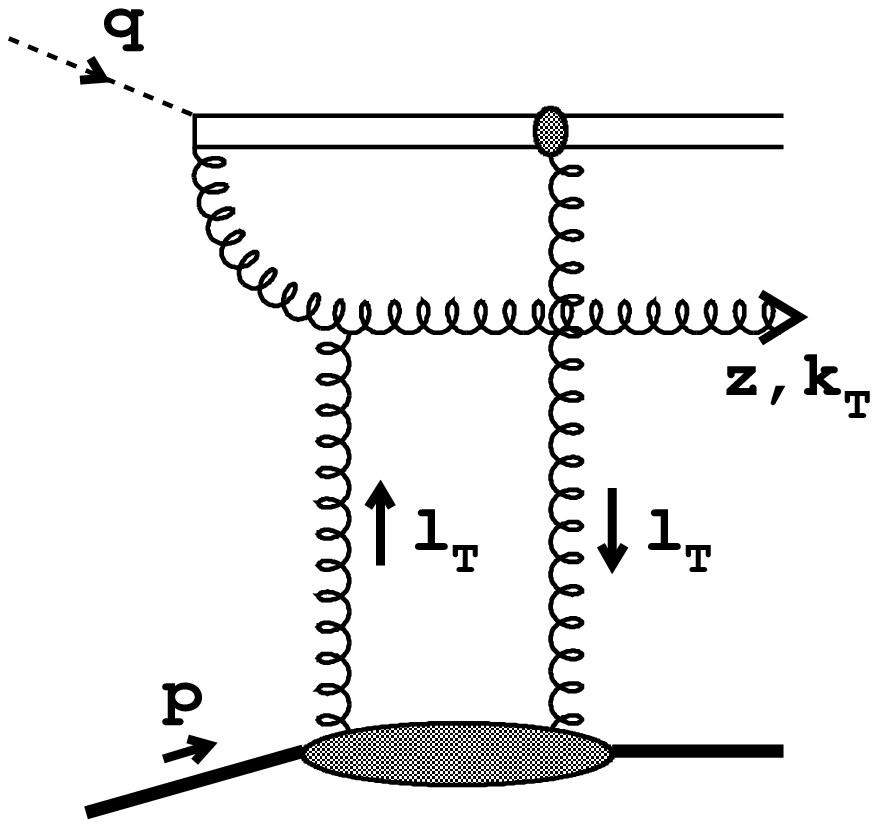,width=6cm}
           }
\vspace*{-0.5cm}
\caption{Diffractive $q\bar{q}$ and $q\bar{q}g$ contributions.
\label{fig:5}}
\end{figure}

DIS diffraction, $\gamma^*p\rightarrow p X$,
is a good test of the parton saturation ideas incorporated
in (\ref{eq:4}). In the simplest case the diffractive system $X$, well separated
in rapidity from the scattered proton, consists of the $q\bar{q}$ pair.
This contribution dominates for the diffractive mass $M^2\sim Q^2$.
In the large diffractive
mass limit, $M^2\gg Q^2$, additional components like $q\bar{q}g$ have to be taken into
account.  These two components are shown in Figure~\ref{fig:5}.

The cross section for the diffractive $q\bar{q}$ production reads \cite{NIKO1}
\be
\label{eq:18}
\frac{d\,\sigma^D_{T,L}}{dt}_{\mid\, t=0}
\,=\,
\frac{1}{16\,\pi}\,
\int d^2\rbo\, dz\,
|\Psi_{T,L}(\rbo,z)|^2\ \hat\sigma^2(x,\rbo),
\ee
where $t$ is the squared momentum transfer to the diffractive system.
Assuming exponential dependence on $t$, $e^{B_D t}$, we divide
(\ref{eq:18}) by the diffractive slope $B_D$ (taken from the experiment)
to obtain the total
cross section $\sigma^D_{T,L}$.  The dipole cross section $\hat\sigma$
is given by the saturation model (\ref{eq:4}) which parameters were determined
by the inclusive structure function analysis. Thus, the description of diffractive
DIS (for $M^2\sim Q^2$) is parameter free and differs from the inclusive one
by the squared dipole cross section.

We can perform a qualitative analysis using the formulae from Section~\ref{sec:3}.
In the DIS case, when $1/Q\ll R_0(x)$, we find  for the diffractive $q\bar{q}$ production
from transverse photons
\be
\label{eq:19}
\sigma^D_T\,\sim\,
\underbrace{\frac{\sigma_0^2}{Q^4 R_0^4}}_{r<2/Q}
~\;+\;~
\underbrace{\frac{\sigma_0^2}{Q^2 R_0^2}}_{2/Q<r<2R_0}
~\;+\;~
\underbrace{\frac{\sigma_0^2}{Q^2 R_0^2}}_{r>2R_0}.
\ee
By the comparison with (\ref{eq:9}),   we see that the leading scaling result
comes from large transverse sizes.  The contribution from $r<2/Q$
is suppressed by the additional power of $Q^2$ (higher twist contribution).
Notice that the DGLAP modification of the dipole cross section
for small $r$ does not influence the leading result. Therefore, DIS diffraction
is strongly sensitive to the region of the transition to saturation in
the dipole cross section (\ref{eq:4}).  In fact,
the saturated form of $\hat\sigma$
is necessary for the observed in the data constant ratio $\sigma^D/\sigma^{tot}$.
Considering the leading contribution to (\ref{eq:9}) and  (\ref{eq:19}), we find
\be
\label{eq:20}
\frac{\sigma^D}{\sigma^{tot}}
\, \sim \,
\frac{1}{\ln (Q^2 R^2_0(x))},
\ee
which is a slowly varying function of $x$ and $Q^2$.

Summarizing the qualitative analysis, by exposing the semi-hard ($2/Q<r<2R_0$)
and non-perturbative  ($r>2R_0$) regions, DIS diffraction is an ideal process
to test parton saturation realized in terms of the dipole cross section (\ref{eq:4}).
The constant ratio $\sigma^D/\sigma^{tot}$ as a function of $x$ and $Q^2$ 
finds a natural explanation in the saturation model.

\begin{figure}[t]
  \vspace*{-0.5cm}
     \centerline{
         \epsfig{figure=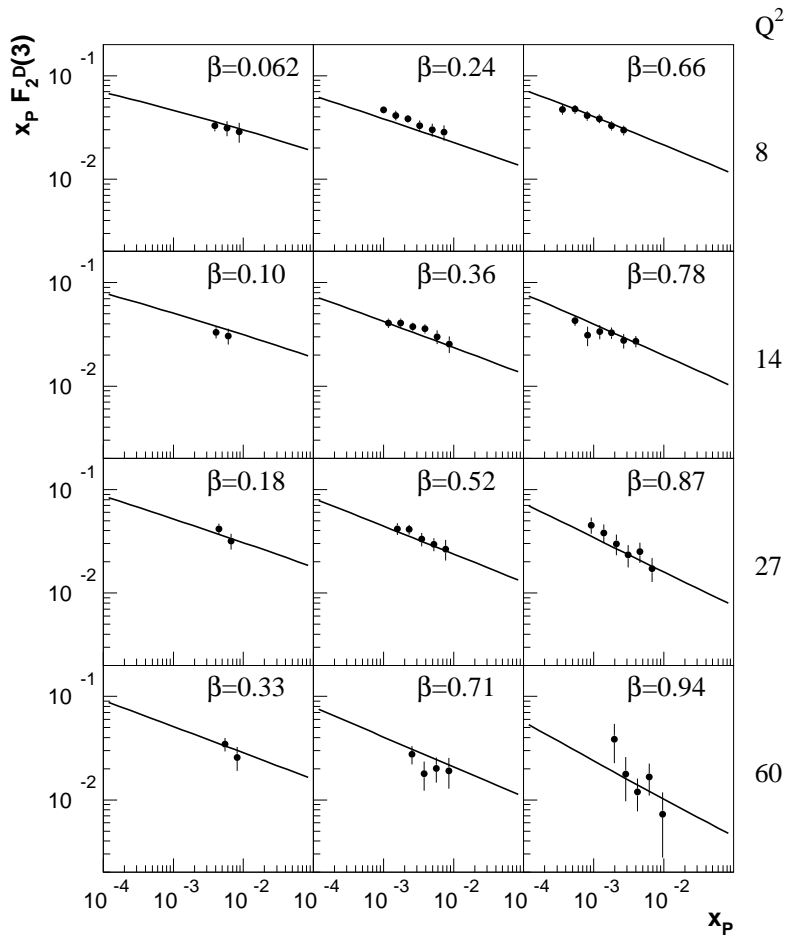,width=11cm}
           }
\vspace*{-0.5cm}
\caption{\it The diffractive structure functions
$F_2^{D(3)}(\beta,Q^2,x_{\funp})$  as a function of $x_{\funp}=(Q^2+M^2)/(Q^2+W^2)$ for
different values  of $\beta=Q^2/(Q^2+M^2)$ and $Q^2$ (in units of $GeV^2$).
The data are from ZEUS \cite{ZEUS99}.
\label{fig:6}}
\end{figure}

In \cite{GBW2} we perform an extensive comparison of the saturation model
predictions with the data by modelling the diffractive state
as shown in Figure~\ref{fig:5}. Good agreement is found
for both the H1 and ZEUS data. In particular, in  Figure~\ref{fig:6} we show the
energy ($\xp$) dependence of the diffractive structure function
which is determined by the energy dependence
found in the inclusive structure function analysis.
In particular, away of the region $M^2\ll Q^2$ ($\beta\rightarrow 1$) where
the higher twist longitudinal $q\bar{q}$ component dominates,  we find for the leading
twist part of the diffractive structure function
\be
\label{eq:21}
F_2^D\,\sim\,\xp^{1-2\alpha_P},
\ee
where the ``effective pomeron intercept''  $\alpha_P=1+\lambda/2\approx 1.15$
($\lambda$ is defined in Section~\ref{sec:2}).
This value is in remarkable agreement with the experimental values:
1.17 by H1 \cite{H197} and 1.13 by ZEUS \cite{ZEUS99}.

We have to add, however, that
the treatment of the $q\bar{q}g$ component,
interacting with the proton with
the same dipole cross section (up to a colour factor) as the $q\bar{q}$ system, goes beyond
the saturation model since this component is not present in the inclusive
analysis. More detailed studies are necessary.

The presented model of DIS diffraction can be extended to include more
complicated diffractive states, e.g. with partons strongly ordered in transverse momenta
(DGLAP configuration). In this case initial diffractive parton distributions are
directly computed from the saturation model
and subsequently evolved using the DGLAP evolution equations.
The results are presented in \cite{GBW3}.

\section{Conclusions}

The idea of parton saturation realized through the proposed model of the dipole cross
section turns out to be very successful in the unified description  of inclusive
and diffractive DIS at small $x$.
In particular, the transition to small $Q^2$ in inclusive DIS,
the constant ratio $\sigma^D/\sigma^{tot}$ and
the energy dependence of diffractive DIS are naturally obtained in the presented
approach. From a formal point of view, these ideas allow to obey
unitarity of the description. The detailed QCD picture of the discussed processes
is still a matter of intensive theoretical studies. From a phenomenological
point of view the most promising is the analysis done in \cite{KOVCH, KOVCH1}
in which multiple pomeron exchanges are responsible for saturation as postulated
in our model. In this case the dipole cross section
(or better the $q\bar{q}$ -- $p$ forward scattering amplitude)
obeys a non-linear evolution equation.
Therefore, the future offers an exciting time for new developments.

\ack
I am grateful to the organizers of the Ringberg Workshop for their kind invitation and for
an excellent organization. I warmly thank  J.~Bartels, H.~Kowalski, M.~W.~Krasny,
J.~Kwieci\'nski,  K.~Peters, S.~Riess, A.~Sta\'sto and particularly  M.~W\"usthoff
for enjoyable collaboration on the subject of this presentation.
I also thank E.~\L obodzinska for careful  reading of the manuscript.
This research has been supported in part by the EU
Fourth Framework Programme
``Training and Mobility of Researchers''  Network,
``Quantum Chromodynamics and the Deep Structure of Elementary
Particles'', contract   FMRX-CT98-0194 (DG~12-MIHT) and by the Polish KBN
grant No.  5 P03B 144 20.  The  Deutsche Forschungsgemeinschaft fellowship is gratefully
acknowledged.


\end{document}